\documentstyle[multicol,aps,epsf]{revtex}
\draft
\begin{document}

\title{Comment on ''Evidence for Coexistence of the Gap and the
Pseudogap in Bi-2212 from Intrinsic Tunneling Spectroscopy"}

\author{M.Ya. Amusia$^{a,b}$ and V.R. Shaginyan$^{c}$
\footnote{E--mail: vrshag@thd.pnpi.spb.ru}}

\address{$^{a\,}$The
Racah Institute of Physics, the Hebrew University, Jerusalem 91904,
Israel;\\ $^{b\,}$Physical-Technical Institute, Russian Academy of
Sciences, 194021 St. Petersburg, Russia;\\ $^{c\,}$ Petersburg Nuclear
Physics Institute, Russian Academy of Sciences, Gatchina, 188350,
Russia}

\maketitle

\begin{abstract}
It is argued that the superconducting and pseudogaps represent
different coexisting phenomena, and this observation speaks against
the precursor-superconductivity scenario of the pseudogap \cite{ky}.
We present brief comments showing that the experimental facts of
paper \cite{ky} do not contradict to observations that the pseudogap
in underdoped Bi-2212 smoothly evolves into a pseudogap above
critical temperature $T_c$ \cite{ding,lsd,norm}, and the large energy
gap below $T_c$ is mainly of superconducting origin \cite{mzo}.  We
show that appearance of this pseudogap is closely related to a new
energy scale observed recently \cite{blk}.  \end{abstract}

\pacs{ PACS numbers: 74.25.Jb, 71.27.+a, 74.20.Fg}

\begin{multicols}{2}

In a recent Letter \cite{ky}, Krasnov {\it at al.}, using the
intrinsic tunneling spectroscopy, observed on small slightly
overdoped and optimally doped Bi$_2$Sr$_2$CaCu$_2$O$_{8+\delta}$
(Bi-2212) mesas that the superconducting gap (SG) does vanish,
while the pseudogap (PG) does not change at $T=T_c$. As a result,
they claim that all this speaks in favor of different origin of
two coexisting phenomena and against the
precursor-superconductivity scenario of the PG. On the other hand,
basing on angle-resolved photoemission spectroscopy (ARPES),
accurate measurements gave firm evidence that in optimally doped
and overdoped samples, a d-wave superconducting gap closes at the
same temperature, $T_c$. However, in underdoped samples, the SG
below $T_c$ smoothly evolves into a PG above $T_c$
\cite{ding,lsd,norm}. Moreover, detailed examination of the
tunneling spectra over wide doping range show  that the large
energy gaps are predominantly of superconducting origin
\cite{mzo}. Thus, the main results of \cite{ding,lsd,norm,mzo} are
in conflict with the ones of \cite{ky}, and we face a number of
contradictions. For instance, we have both negative and positive
answers to at least the following statements: 1) the PG exists in
optimally doped samples; 2) there is no direct relation between the
SG and the PG; 3) the SG does not smoothly evolve into the PG; and
4) the energy gap at $T\leq T_c$ is composed from the SG and PG. In
our Comment we show that these contradictions are only visible and
explain the origin of PG described in \cite{ky}.

For the readers convenience, we summarize here the main
experimental results of \cite{ky}.

At low $T$, the data show a sharp peak in the dynamic conductance
$\sigma(v)$, which is attributed to the SG voltage,
$v_s=2\Delta_s/e$ (our notations correspond to that in \cite{ky}).
With increasing $T$, the peak at $v_s$ reduces in amplitude and
shifts to lower voltages, reflecting the decrease in
$\Delta_s(T)$, while $2\Delta_s(T=0)\simeq 66$ meV. At $T\simeq
83$ K ($<T_C\simeq 93$ K) the superconducting peak is smeared out
completely and a hump in conductance at $v=v_{pg}\simeq 70$ meV
remains. Then, there are no sharp changes at $T_c$. The PG
persists in the superconducting state. At $T<T_c$ with decreasing
temperature, the superconducting peak shifts to higher voltages,
increases in amplitude, and eventually the PG hump is washed out
by much stronger superconducting peak. For optimally doped mesas,
the PG hump can be resolved at $T>60$ K, i.e., well below $T_c$.

Now we turn to a consideration of the cited above experimental
results related to the peak at $v_{pg}$. Recently a new energy
scale for quasiparticle dispersion in superconducting and normal
states of Bi-2212 was discovered, which manifests itself as a
break in the quasiparticle dispersion near $50\pm 15$ meV binding
energy $E_0$ \cite{blk} that results in a change in the
quasiparticle velocity. An explanation of these experimental
findings was given in \cite{ams,ars} which relates the scale $E_0$
to the SG, $E_0\simeq 2\Delta_1$, with $\Delta_1$ being the
maximum superconducting gap at $T=0$. The appearance of this scale
is a consequence of the fermion condensation: a quantum phase
transition which can take place in high temperature
superconductors \cite{ams1}. At $T\leq T_c\ll T_f$ the scale
$E_0$ is approximately temperature independent. At $T_c\leq T\ll
T_f$, we have $E_0\simeq 4T$ if there is no pseudogap which
evolves from the SG as discussed in \cite{ding,lsd,norm}.
Otherwise $T_c$ is substituted by the temperature $T^*$ above
which the pseudogap closes. Here $T_f$ is the temperature, about
which the fermion condensation effects vanish. Thus, the
quasiparticle dispersion can be presented by two straight lines,
characterized by effective masses $M_{FC}$ and $M_L$, and
intersecting near the binding energy $E_0$.  The slower dispersing
low energy part is defined by $M_{FC}$, while $M_L$ defines the
faster dispersing high energy part. In the region where the SG
reaches its maximum value $\Delta_1$ the ratio $M_L/M_{FC}\ll 1$
leading to a strong change in the density of states at $E_0\simeq
2\Delta_1$. We can now conclude that at $T_c\leq T$, it is the
scale $E_0$ that produces a hump in conductance at $v=v_s\simeq
70$ meV $\simeq 2\Delta_1$. At $T\leq T_c$, the superconducting
peak emerges. With decreasing temperature it moves to higher
energies until it reaches its maximum value $2\Delta_1$ merging
with the scale $E_0$.

In conclusion, we see that PG observed in \cite{ky} is related to
the scale $E_0$ and therefore this PG has a nature different from
the one described in \cite{ding,lsd,norm}. This PG does not
contradict to statement that SG does smoothly evolve into the PG,
and there is no grounds to think that the energy gap at $T\leq
T_c$ is composed from the SG and PG. We are led to a conclusion
that the main results of papers \cite{ding,lsd,norm,mzo} are not
at variance with experimental findings of \cite{ky}. The main
result of \cite{ky} can be interpreted as an independent
confirmation of the existence of a new energy scale at
temperatures below and above $T_c$ which was firstly observed in
\cite{blk}.

\end{multicols}

\end{document}